\documentclass[twocolumn, amsmath, prl, superscriptaddress]{revtex4-1}
\usepackage{xcolor, mathtools}
\usepackage[hidelinks]{hyperref}
\begin{document}

\title{Monofractality in the solar wind at electron scales: insights from KAW turbulence}

\author{Vincent \textsc{David}}
\affiliation{Laboratoire de Physique des Plasmas, \'Ecole polytechnique, F-91128 Palaiseau Cedex, France}
\affiliation{Universit\'e Paris-Saclay, IPP, CNRS, Observatoire Paris-Meudon, France}

\author{S\'ebastien \textsc{Galtier}}
\affiliation{Laboratoire de Physique des Plasmas, \'Ecole polytechnique, F-91128 Palaiseau Cedex, France}
\affiliation{Universit\'e Paris-Saclay, IPP, CNRS, Observatoire Paris-Meudon, France}

\author{Romain \textsc{Meyrand}}
\affiliation{Department of Physics, University of Otago, 730 Cumberland St., Dunedin 9016, New Zealand}

\begin{abstract}
The breakdown of scale invariance in turbulent flows, known as multifractal scaling, is considered a cornerstone of turbulence.
In solar wind turbulence, a monofractal behavior can be observed at electron scales, in contrast to larger scales where multifractality always prevails. Why scale invariance appears at electron scales is a challenging theoretical puzzle with important implications for understanding solar wind heating and acceleration.
We investigate this long-standing problem using direct numerical simulations (DNS) of three-dimensional electron reduced magnetohydrodynamics (ERMHD).
Both weak and strong kinetic Alfvén waves (KAW) turbulence regimes are studied in the balanced case.
After recovering the expected theoretical predictions for the magnetic spectra, a higher-order multiscale statistical analysis is performed.
This study reveals a striking difference between the two regimes, with the emergence of monofractality only in weak turbulence, whereas strong turbulence is multifractal.
This result, combined with recent studies, shows the relevance of collisionless weak KAW turbulence to describe the solar wind at electron scales. 
\end{abstract}

\maketitle

\paragraph{Introduction. \label{sec:Intro}}

After its discovery in the 1950s \citep{Parker_1958}, the solar wind was extensively explored using \textit{in situ} measurements, giving the interplanetary medium a unique  position in astrophysics. 
These studies have significantly advanced our understanding of space plasmas at (sub-MHD) electron scales, where ion and electron heating mechanisms remain elusive in a turbulent, collisionless medium.
Magnetic fluctuations at electron scales were detected in the early 1980s \citep{Denskat_1983}, but it was with Cluster mission that routine observation of magnetic fluctuations in a new range of frequencies was firmly established, the origin of which can be attributed to turbulence \citep{Alexandrova_2008, Sahraoui_2009}.
For this frequency range, a second breakthrough came with the discovery of a monofractal behavior \citep{Kiyani_2009}, subsequently confirmed by other studies \citep{Chhiber_2021}.
This contrasts sharply with the observed multifractal behavior, at (MHD) low-frequencies in the solar wind \citep{Kiyani_2009}, as well as in many turbulent systems \cite{Frisch_1995}.
A third important step was recently taken with the finding of a correlation between the weakening of magnetic fluctuations and the presence of ion cyclotron waves \citep{Bowen_2023}.
These observations seem to support the existence of a helicity barrier at the ion gyroradius scale \citep{Meyrand_2021} where highly Alfv\'enic imbalanced turbulence is prevented from cascading, leading to ion-cyclotron resonant heating \cite{Squire_2022} and, subsequently, to the emergence of weak balanced KAW turbulence at electron scales \citep{Bowen_2022}.

Over the past two decades, a theoretical picture of the kinetic turbulent cascade has been proposed in the presence of a relatively strong mean magnetic field (in velocity units) $\boldsymbol{b_0} = b_0 \hat{\boldsymbol{e}}_z$ ($\vert \hat{\boldsymbol{e}}_z \vert = 1$), in which a cascade develops from MHD to electron scales \citep{Schekochihin_2009}.
A fundamental ingredient of the phenomenology used is the critical balance (CB) assumption \citep{Goldreich_1995}, which states that a scale-by-scale equilibrium is established between linear ($\tau_\mathrm{lin}$) and nonlinear ($\tau_\mathrm{nl}$) times.
This phenomenology of strong turbulence applied to ERMHD, a fluid model containing KAW, leads to anisotropy, with a cascade mainly in the direction perpendicular ($\perp$) to $\boldsymbol{b_0}$, and a magnetic spectrum in $k_\perp^{-7/3}$  \citep{Biskamp_1996,Cho_2004,Schekochihin_2009}, with $k_\perp = \sqrt{k_x^2 + k_y^2}$ and $k_j$ the $j=\{x,y,z\}$ component of the wavevector $\boldsymbol{k}$.
However the scaling discrepancy, often close to $-8/3$ in solar wind observations \citep{Huang_2021}, has been linked to Landau damping \citep{Howes_2011, Tenbarge2013, Groselj_2019}.
A phenomenology for strong KAW turbulence has also been proposed to predict the $-8/3$ index \citep{Boldyrev_2012}.
The problem with these theoretical studies is that they mostly ignore the monofractal behavior mentioned above.
This property is generally not observed in strong turbulence, which is multifractal \citep{Cho_2009,Zhou_2023b}, but can be in weak turbulence \citep{Galtier_2023b}. 
Such a theory was developed for incompressible Hall MHD \citep{Galtier_2006}, a system that reduces to electron MHD at electron scales (with whistler waves) \citep{Galtier_2003b}.
The model was criticized because compressibility is a fundamental ingredient of KAW, but as shown later \citep{Galtier_2015,Schekochihin_2009}, in the presence of a strong $b_0$, a simple rescaling allows us to go from ERMHD to  electron MHD.
Therefore, the physics of weak turbulence is the same for both systems. 
Interestingly, it was found that $k_\perp^{-8/3}$ is a solution of collisionless weak KAW turbulence \citep{David_2019}. 

In this Letter, we investigate KAW turbulence in the weak and strong regimes.
Our numerical study reveals that only the former is monofractal, suggesting that the solar wind at electron scales can be governed by collisionless weak KAW turbulence.

\paragraph{Theory. \label{sec:Theory}}
ERMHD will be used to study KAW turbulence.
This system can be derived from a gyrokinetic or a fluid approach, in  the limit of strong anisotropy $k_\| \ll k_\perp$ and massless isothermal electrons \citep{Schekochihin_2009,Boldyrev_2013,Galtier_2023b}.
It is valid for $k_\perp \rho_i > 1$ and $k_\perp \rho_e < 1$, where $\rho_{i,e}$ are the Larmor radius of ions/electrons; ERMHD reads
\begin{equation} \label{eq1}
    \partial_t \psi = -\frac{d_i b_0^2}{\kappa-1} \nabla_\| n , \quad
    \partial_t n = \frac{d_i (\kappa-1)}{\kappa} \nabla_\| \nabla_\perp^2 \psi,
\end{equation}
with $\kappa = 1+2/(\beta_i (1+Z/\tau))$ the compressible coefficient,  $\beta_i$ the ratio between thermal and magnetic pressures for ions, while $\tau \equiv T_i/T_e$ and $Z \equiv q_i/e$ describe the ion/electron temperature and charge ratios, respectively.
$n$ denotes the relative (to the mean value $n_0$) electron density perturbation (with $n \ll n_0$) and is linked to the $z$-component of the magnetic field by the pressure balance condition $n = \left( 1 - \kappa \right) b_z/b_0$ \citep{Schekochihin_2009}.
$d_i$ is the ion inertial length, and $\psi$ the magnetic stream function ($\boldsymbol{b}_\perp = \hat{\boldsymbol{e}}_z\times \boldsymbol{\nabla}_\perp \psi$).
By definition,  $\nabla_\perp^2 \equiv \partial_{xx} + \partial_{yy}$, and $\nabla_\| \cdot \equiv \partial_z \cdot + b_0^{-1} \left\{ \psi, \cdot \right\}$, where the linear part denotes the gradients along $\boldsymbol{b_0}$ and the nonlinear part (i.e., $\{ \psi, g \} = \partial_x \psi \partial_y g - \partial_y \psi \partial_x g$, with $g$ a scalar field) stands for the gradients along the local field $\boldsymbol{b_0} + \boldsymbol{b}$ (with $\vert \boldsymbol{b} \vert \ll b_0$).
By performing an elementary change of variable ($\psi \to \psi \, \kappa / \left( \kappa-1 \right) $ and $b_0 \to b_0 \sqrt{\kappa}$), our results become valid regardless the value of the compressible coefficient $\kappa$.
In the linear case, KAW are solutions of Eq. (\ref{eq1}); these waves are oblique, dispersive and adhere to $\omega = d_i b_0 k_\perp k_\|$ in the anisotropic limit \citep{Galtier_2023b}. Note that ERMHD conserves energy $E \equiv \int ( \left\vert \boldsymbol{\nabla}_\perp \psi \right\vert^2  + b_z^2 ) \mathrm{d}\boldsymbol{x}$, and helicity $H \equiv \int \psi b_z \mathrm{d}\boldsymbol{x}$.

Weak and strong wave turbulence can be differentiated through the parameter
\begin{equation} \label{eq:chi}
    \chi \left( k_\perp, k_\| \right) \equiv \frac{\tau_\mathrm{lin}}{\tau_\mathrm{nl}} \simeq \sqrt{\frac{2 k_\perp^3 E \left( k_\perp, k_\| \right)}{k_\| b_0^2}},
\end{equation}
where $\tau_\mathrm{lin} \simeq 1/ \omega$, $\tau_\mathrm{nl} \simeq 1 / (d_i k_\perp^2 b_k)$, and $k_\| \simeq k_z$ which is less precise in the strong regime.
Here, $b_k \simeq \sqrt{2 k_\perp k_\| E(k_\perp, k_\|)}$ represents the amplitude of the magnetic field and $E(k_\perp, k_\|)$ is the axisymmetric bidimensional magnetic energy spectrum.
The weak regime corresponds to $\chi \ll 1$ for all wavenumbers, and the cascade is seen as the result of numerous collisions between (mainly) counter-propagating wave packets. This is basically a multiple timescale problem involving three-wave interactions, and for which a natural asymptotic closure  exists \cite{Benney_1966,Newell_2001,Nazarenko_2011b}.
An exact solution to the problem is the magnetic energy spectrum $E(k_\perp, k_\|) \propto k_\perp^{-5/2} k_\|^{-1/2}$ \cite{Galtier_2003b,Galtier_2023b}.
According to CB phenomenology, the strong regime corresponds to $\chi \sim 1$ \cite{Goldreich_1995}.
In this case, there is no rigorous spectral theory and DNS is used to explore this regime \cite{Biskamp_1996, Biskamp_1999, Cho_2009, Cho_2011, Kim_2015}.
The phenomenology suggests a magnetic energy spectrum $E(k_\perp, k_\|) \propto k_\perp^{-7/3} k_\|^{-1}$ \cite{Biskamp_1999, Schekochihin_2009} and a scaling relation $k_\| \propto k_\perp^{1/3}$ \cite{Cho_2004}.
The last regime $\chi \gg 1$ is neither relevant nor sustainable due to the causal impossibility of maintaining $\tau_\mathrm{nl} \ll \tau_\mathrm{lin}$ \cite{Schekochihin_2009, Schekochihin_2022}.

\paragraph{Numerical Setup. \label{sec:Numerics}}
DNS are performed with \texttt{AsteriX} \cite{Meyrand_2019, Meyrand_2021}, a modified version of the pseudo-spectral code \texttt{TURBO} \cite{Teaca_2009}.
Time stepping is done using a third-order modified Williamson algorithm \cite{Williamson_1980}.
A triple periodic cubic box of size $L=2\pi$ is used with $N_\perp^2 \times N_z$ Fourier modes.
Simulations use a recursive refinement technique \cite{Meyrand_2021}, with the highest resolution being $N_\perp = 1024$ and $N_z = 128$, from which we develop our analysis.
To address aliasing effects in nonlinear terms, a phase shift method is employed, resulting in partial dealiasing \cite{Patterson_1971}.

The solved system, including additional dissipative (hyper-Laplacian operator) and forcing terms, corresponds to the normalized ($b_0 = d_i = 1$) equations
\begin{subequations}
\begin{eqnarray}
    {\partial_t \psi} &=& - \nabla_\| n + \eta \nabla^6 \psi + f^\psi,\\
    \label{eq:bz-num}
     \partial_t n &=& \nabla_\| \nabla_\perp^2 \psi+ \eta \nabla^6 n + f^n,
\end{eqnarray}
\end{subequations}
where $\eta$ is a dissipative coefficient.
Two DNS of balanced KAW turbulence are conducted to study the weak and strong regimes.
The only difference between these simulations is the constant energy injection rate, with  $\varepsilon=10^{-3}$ in the first and $\varepsilon=0.5$ in the second.
To reach a stationary state, $\varepsilon$ is controlled \textit{via} forcing terms ($f^\psi$ and $f^n$) with random phases, selectively applied at $\left\vert k_z \right\vert = 1$ in the range $1.5 < k_\perp < 2.5$ (large-scale fluctuations with Gaussian distributions).
These terms allow a precise control of energy and helicity (taken to be zero -- see \cite{Passot_2019, Miloshevich_2020} for the case with helicity), promoting chaotic motions necessary for turbulent behavior. 
\begin{figure}
    \centering
    \includegraphics[width=\columnwidth]{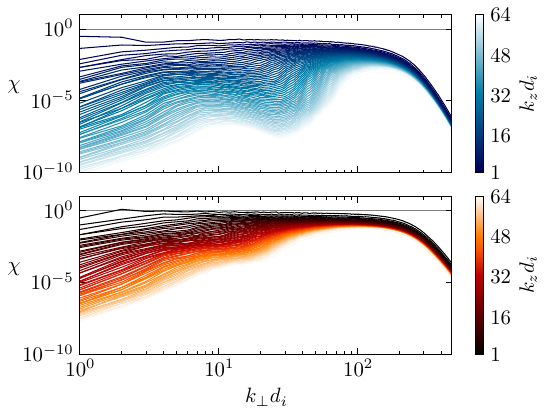}
    \caption{Variation of $\chi$ in the weak (top) and strong (bottom) regimes. The gray horizontal lines display $\chi=1$ and the intensity of the colors corresponds to different values of $k_z$.}
    \label{fig:chi}
\end{figure}

\paragraph{Results. \label{sec:Validation}}
To confirm the presence of the desired regimes, we first examine $\chi$. 
Figure \ref{fig:chi} demonstrates that both DNS satisfy the expected criterion.
In the weak regime, $\chi \ll 1$ is observed across all wavenumbers.
In the strong regime, the low parallel modes display $\chi \sim 1$.
However, as we move to higher parallel modes, $\chi$ decreases, deviating from the CB hypothesis.
Although some modes belong to the weak turbulence regime (specifically for $k_z > 32$), their contribution to the dynamics is negligible as they are energetically subdominant by several orders of magnitude.
Therefore, this DNS is mainly governed by strong turbulence.

\begin{figure}
    \centering
\includegraphics[width=\columnwidth]{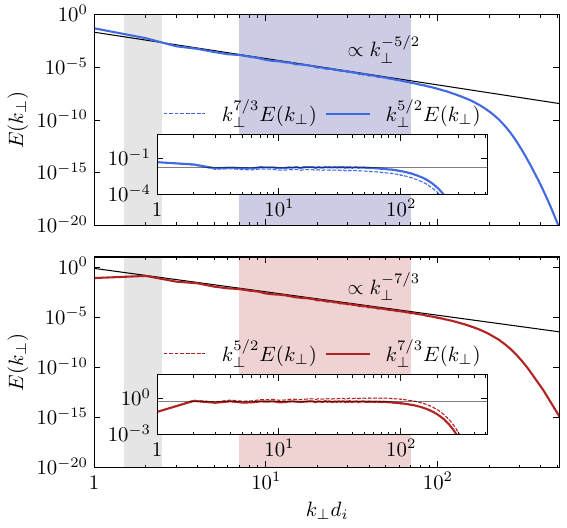}
    \caption{Transverse energy spectra for the weak (top) and strong (bottom) regimes. The gray columns show the forcing scales, while the blue and red columns indicate the regions of the inertial range where the intermittency is studied (see Figure \ref{fig:struc-func}). Insets provide compensated spectra by the theoretically expected scaling (solid and dashed lines).}
    \label{fig:spectra}
\end{figure}

Since the energy transfer is weaker along the mean magnetic field, we will use fewer Fourier modes in the $z$ direction and will thus primarily focus on the perpendicular dynamics.
 Figure \ref{fig:spectra} displays the one-dimensional axisymmetric transverse magnetic spectra $E(k_\perp)$ for both regimes.
 In line with theoretical predictions, the spectra obtained by integrating on a cylinder aligned with $\boldsymbol{b_0}$ exhibit power law indices of approximately -5/2 and -7/3 in the weak and strong regimes, respectively, over more than a decade.

\begin{figure}
    \centering
    \includegraphics[width=\columnwidth]{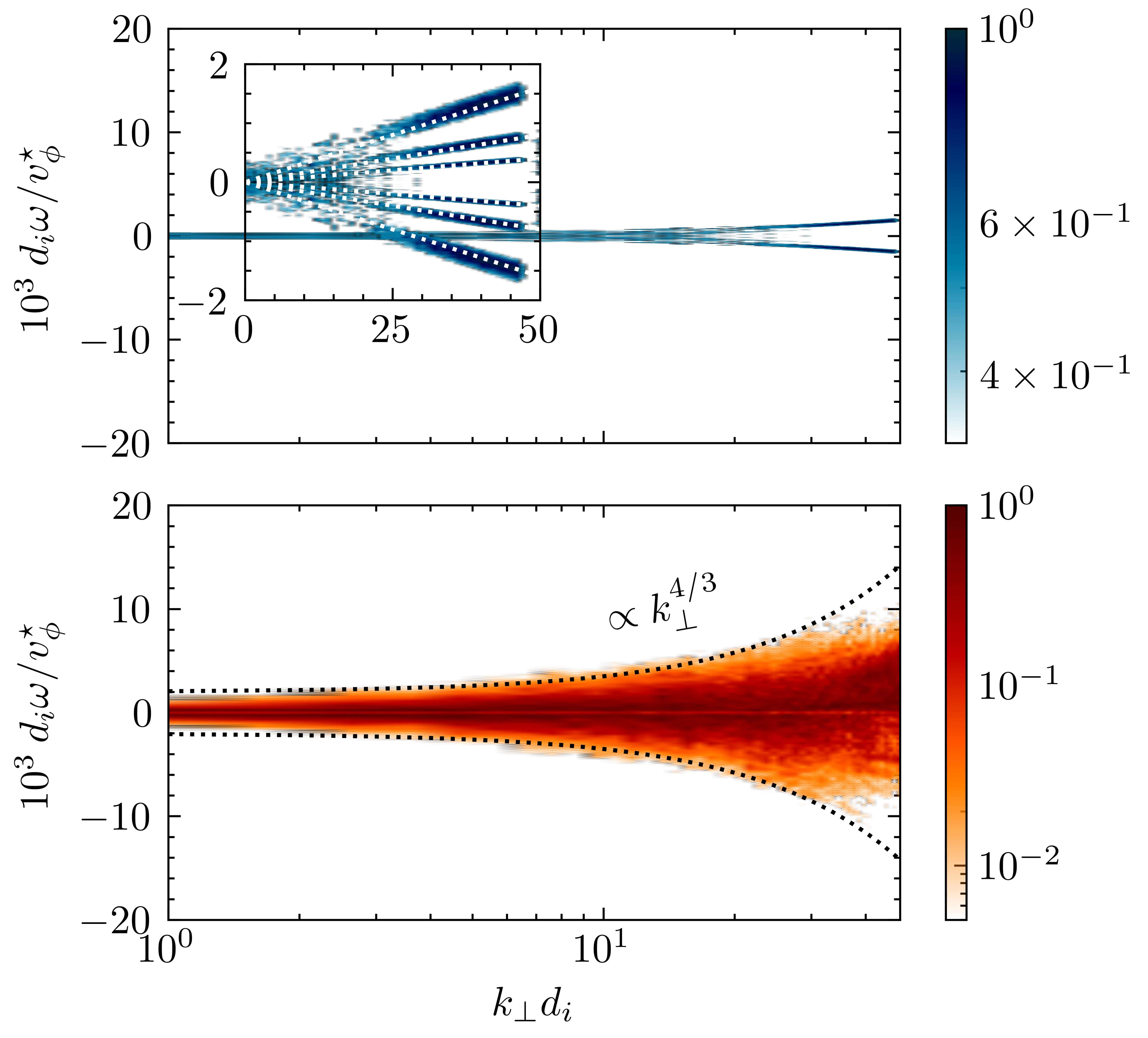}
    \caption{Normalized space-time energy spectra in the weak (top) and strong (bottom) regimes, at $k_z=4$ (with $v_\phi^\star \equiv d_i b_0 k_\perp^\star$ and $k_\perp^\star = \sqrt{2}$). Inset (top): spectra in linear scales for $k_z=\{\pm 1,\pm 2, \pm 4\}$ with corresponding theoretical dispersion relations for KAWs (dotted lines). Bottom: the power law $\pm k_\perp^{4/3}$ (dotted lines) is shown for comparison with the CB law.}
    \label{fig:kw}
\end{figure}

The wavenumber-frequency spectrum $E(k_\perp,k_z,\omega)$ provides further evidence of the difference between weak and strong KAW turbulence. It is also the best way to demonstrate the presence of weak turbulence. 
These spectra are constructed by following the temporal variations (over several linear wave periods) of fluctuating fields in Fourier space along the diagonal $k_x = k_y$ (at a fixed $k_z$). 
Figure \ref{fig:kw} shows these spectra with an inset superposing $k_z=\{\pm 1, \pm 2, \pm 4\}$. In the weak regime, wave interactions become evident with the emergence of dispersion relations of KAW. 
Note that signals with low $\omega$ are mainly limited by time integration: the lower the frequencies, the longer the signal must be integrated.
In contrast, the strong regime shows no discernible patterns, with a broad range of frequencies excited in a region delimited approximately by the power laws $\pm k_\perp^{4/3}$.
Interestingly, this corresponds to CB phenomenology.
This region contains the KAW dispersion relations (at $k_z= \pm 4$), which means that nonlinearities are strong enough to include non-resonant three-wave interactions and thus considerably broaden the two branches but remain limited by the $\chi \sim 1$ condition (a similar situation was found in MHD turbulence \citep{Meyrand_2016}).
Therefore, we see that strong wave turbulence does not excite fluctuations corresponding to $\chi \gg 1$.

\begin{figure}
    \centering
    \includegraphics[width=\columnwidth]{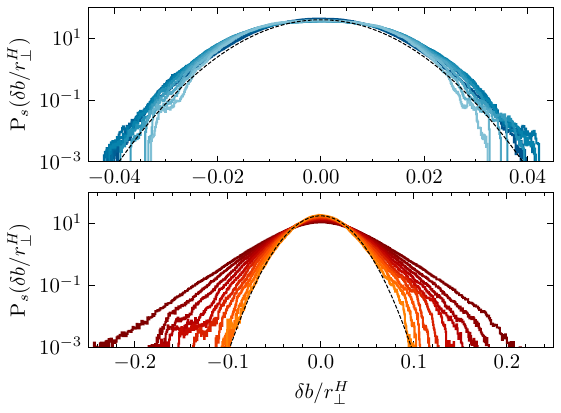}
    \caption{PDFs of the magnetic field increments $\delta b$ as a function of 
    $r_\perp/L$, ranging from dark to light colors in the interval $\left[1.38, 13.8\right] \times 10^{-2}$
    for the weak (top) and strong (bottom) regimes. The black dashed lines correspond to the Gaussian fit of the PDFs.}
    \label{fig:histo}
\end{figure}

Intermittency is assessed through the departure from Gaussian behavior in the probability density function (PDF) of the magnetic field modulus increment $\delta b = b(\boldsymbol{x} + \boldsymbol{r}) - b(\boldsymbol{x})$.
We focus on increments within the $N_z$ perpendicular planes. Figure \ref{fig:histo} shows that in the weak turbulence regime the PDFs are Gaussian-like distributions with negligible tails.
On the other hand, the strong turbulence regime exhibits significant non-Gaussian tails with decreasing perpendicular increment distance $r_\perp$. 
Additionally, we employed a rescaling operation \cite{Kiyani_2006} to normalize the increments 
$P_s ( \delta b / r_\perp^H ) = r_\perp^H P ( \delta b, r_\perp )$, where $P_s$ represents the rescaled PDF and $P$ denotes the original PDF.
The value of $H$ is determined through a fitting analysis.
A remarkable outcome is the collapse of the different PDFs in the weak turbulence regime when $H=0.75$. 
This serves as validation for the self-similar characteristics of this regime and is consistent with the power law index -5/2.
It is interesting to note that this value of $H$ aligns with solar wind observations where $H \sim 0.8$ \cite{Kiyani_2009}, and that the properties found here with respect to (non) Gaussianity are compatible with observations made near the Sun \citep{Bowen_2023}.

\begin{figure}
    \centering
    \includegraphics[width=\columnwidth]{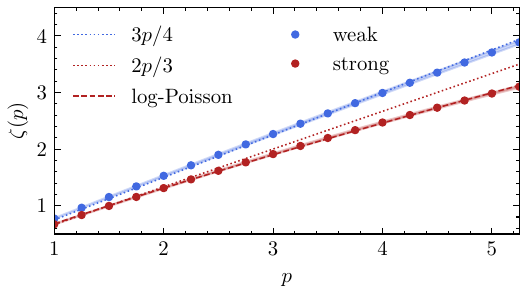}
    \caption{Scaling exponents $\zeta (p)$ of structure functions in the weak (blue line) and strong (red line) regimes. Blue and red dotted lines indicate expected self-similar profiles (see text), while shaded areas represent fitting errors associated with the exponents. The blue and red disks represent the computed values of $\zeta(p)$, and the red dashed line corresponds to a log-Poisson model.}
    \label{fig:struc-func}
\end{figure}

We now introduce the $p$-order structure functions as $S_p \equiv \left\langle \left\vert \delta b \right\vert^p \right\rangle = C_p r^{\zeta(p)}$, where $\left\langle \cdot \right\rangle$ represents the ensemble average, $\zeta(p)$ is the scaling exponents measured in the inertial range, and the coefficients $C_p$ are constants.
Examining higher-order structure functions provides a means to investigate increasingly smaller scales.
As $p$ increases, the structure function becomes more sensitive to fine-scale gradients, enabling the identification of rare events within the PDFs.
For the weak and strong regimes to align with the magnetic energy spectra, they must satisfy $\zeta(2) = 3/2$ and $\zeta(2) = 4/3$, respectively.
Assuming self-similarity, we can derive the scaling exponents as $\zeta(p) = 3p/4$ and $\zeta(p) = 2p/3$ for the weak and strong regimes, respectively.
In order to further explore this topic, we compute the $S_p$ values for quarter integers $p \in [1,5]$, considering the limitations imposed by the available data points \cite{DudokDeWit_2004}.
As with the PDF analysis, increments are calculated in perpendicular planes.
We took a dozen equidistant planes in the $z$ direction for each of the two times considered. 
This process yielded a total of $\sim 10^9$ samples for each value of $p$.
Figure \ref{fig:struc-func} displays the scaling exponents $\zeta(p)$. A distinct linear relationship emerges following the self-similarity prediction for weak wave turbulence. 
In the strong turbulence regime, a departure from self-similar scaling becomes evident as $p$ increases, displaying a multifractal nature. 
This behavior is consistent with a phenomenological log-Poisson law \cite{She_1994,Biskamp_2003,Meyrand_2015}
\begin{equation} \label{eq:poisson}
\zeta(p) = \frac{2p}{3}(1-\Delta)+C_{0}-C_{0}\left(1 - \frac{\Delta}{C_0}\right)^{2p/3},
\end{equation}
where $\Delta=1/3$ (value compatible with a $-7/3$ spectrum) and $C_0=1.1$ is the fractal co-dimension determined empirically (for two-dimensional dissipative structures we expect $C_0=1$).
Note that for deriving expression (\ref{eq:poisson}) we have used the relation $S_p \sim \langle \varepsilon^{2p/3} \rangle r_\perp^{2p/3}$, with $\langle \varepsilon \rangle$ the mean rate of energy dissipation.
These observations suggest that, in contrast to the strong regime that concentrates energy in sparse plasma regions to develop coherent structures, the weak regime exhibits a more even energy distribution throughout the plasma.
This aligns with the absence of strong nonlinearities, the lack of distinct structures, and the random phase approximation.

\paragraph{Conclusion. \label{sec:Conclusion}}

Our main results are as follows:
(i) for the first time, the weak regime is produced with a spectral behavior in agreement with the theory and, in particular, with the analytical solution \citep{Galtier_2003b,Galtier_2023b}, proving that this solution is attractive;
(ii) spectral properties in the strong regime align with the CB phenomenology;
(iii) intermittency properties of KAW turbulence are different depending on the regime, with a standard multifractality in strong turbulence and a monofractality in weak turbulence.
These features are quite different from those found at MHD scales, where multifractality is observed numerically for both regimes \citep{Muller_2000,Meyrand_2015}.
Note that a monofractal behavior has already been observed for inertial wave turbulence \citep{vanBokhoven_2009,Mininni_2010}, a regime with similar properties \citep{Galtier_2020}.

A striking feature of solar wind turbulence is this monofractal scaling observed at electron scales \citep{Kiyani_2009,Chen_2014, Kiyani_2015, Alberti_2019, Chhiber_2021, Gomes_2023}.
While the CB phenomenology has received significant attention, the intermittent aspect has been essentially overlooked in numerical simulations devoted to the solar wind \citep{Howes_2011,Tenbarge2013,Groselj_2019} (see, however, \citep{Zhou_2023b}).
Here, we have shown that weak KAW turbulence can reproduce this previously elusive monofractal behavior.
This regime is also characterized by a PDF close to a Gaussian, with negligible non-Gaussian wings.
It is plausible that these wings will be greater if the Reynolds number is higher or the statistics are improved, but the most recent observations made near the Sun with the PSP mission \citep{Bowen_2023} reveal that fluctuations at electron scales are often characterized by a PDF close to Gaussianity.
Therefore, the present work provides a solid interpretation of these observations.
We have also shown that the strong regime has a multifractal behavior compatible with a phenomenological log-Poisson law. This regime is relevant for the solar wind when turbulence is balanced \citep{Bowen_2023}.

The stationary solution for weak KAW turbulence is a power-law energy spectrum with an index $-5/2$, whereas observations often show $-8/3$ \citep{Alexandrova_2012,Bale_2019,Huang_2021,Bowen_2023}.
Kinetic effects can produce a steepening of the spectrum \cite{Howes_2011}, however, it has recently been realized that, assuming highly local nonlinear interactions, $-8/3$ is an attractive solution for collisionless KAW turbulence \cite{David_2019}. 
In the absence of collision (ie., viscous-type term), the bounce of the spectrum  observed when the cascade reaches `viscous' scales cannot exist, and the self-similar solution not predictable by phenomenology should be preserved.

\acknowledgements
We acknowledge the IDCS at the École polytechnique for providing us with computer resources.
Part of this work was supported by a grant from the Simons Foundation (Grant No. 651461, PPC). 

\bibliography{biblio}
\end{document}